\documentclass[12pt,A4]{article}
\usepackage{array}
\usepackage{epsfig}
\usepackage{amsmath}
\newcommand{\Chi}{\mathrm X}
\newcommand{\be}{\begin{equation}}
\newcommand{\ee}{\end{equation}}
\newcommand{\bea}{\begin{eqnarray}}
\newcommand{\eea}{\end{eqnarray}}
\newcommand{\m}{\mathbf}
\def\la{\mathrel{\mathpalette\fun <}}

\def\fun#1#2{\lower3.6pt\vbox{\baselineskip0pt\lineskip.9pt
\ialign{$\mathsurround=0pt#1\hfil##\hfil$\crcr#2\crcr\sim\crcr}}}

\begin{document}
\title{RIGHT-LEFT ASYMMETRY OF RADIATION FROM FISSION}
\author{F.F.Karpeshin \\
\em Fock Institute of Physics, St. Petersburg State University \\
RU-198504 St. Petersburg, Russia}

\maketitle

\begin{abstract}
The effect of the right-left asymmetry is considered
in the angular distribution of gamma quanta from fission of $^{235}$U
by polarised thermal neutrons, which depends on the polarisation
of the neutrons with respect to the gamma---fission plane.
Electric dipole radiation from fission fragments arising due to the
Strutinsky---Denisov induced polarisation mechanism may give rise to such
an effect.   Earlier, this
mechanism was shown to fit the non-statistical part observed in the $\gamma$
spectrum from spontaneous fission of $^{252}$Cf.
The calculated value of the magnitude of the asymmetry parameter
is on the level of  10$^{-4}$.
That is in agreement with the current experimental data.  A crucial
experiment to give a more definite picture of the concrete mechanism
would be determination  of the energy of the quanta responsible for the
asymmetry. Detection of the quanta with the energy of $\sim$~5~MeV
approaching the GDR is needed in order to identify prompt
gamma rays emitted at the stage of fissioning.
\end{abstract}

{\bf PACS: 24.80.+y,   24.75.+i,    24.70.+s}

{\bf Key words: }  Fission, Right-left asymmetry in radiation from
fission

\newpage
\section{Introduction}

First results of the current experiment on search for the angular
correlation in the emission of gamma quanta from fission
of $^{235}$U induced by thermal polarised neutrons were surprising
and of great interest \cite{dan}. Polarisation of the
neutron beam defines the natural quantization axis $z$
in the laboratory frame. In ref. \cite{dan} fission fragments and
gammas were detected in the orthogonal geometry, that is in the
($x$, $y$) plane, which is perpendicular to the neutron polarisation
(Fig. 1). Let $y$ be the fission axis. In \cite{dan},
$\gamma$ quanta were detected under the angle of $\vartheta_{exp}$
with respect to the fission axis, that is axis of $y$
in the frame presented in Fig. 1. In the ($x$, $y$) plane,
angle $\vartheta_{exp}$ is complimentary
to the azimuth angle $\phi$ in Fig. 1:
\be
\vartheta_{exp}=\frac{\pi}2 - \phi\;.   \label{theta}
\ee
For this reason, we will refer the results to the angle $\vartheta_{exp}$,
using (\ref{theta}). There is obtained
an evidence that the $\gamma$ emission probability depends
on the right or left  neutron polarisation with respect to the ($x$, $y$)
plane.
\begin{figure}[!b]
\centerline{ \epsfxsize=11cm\epsfbox{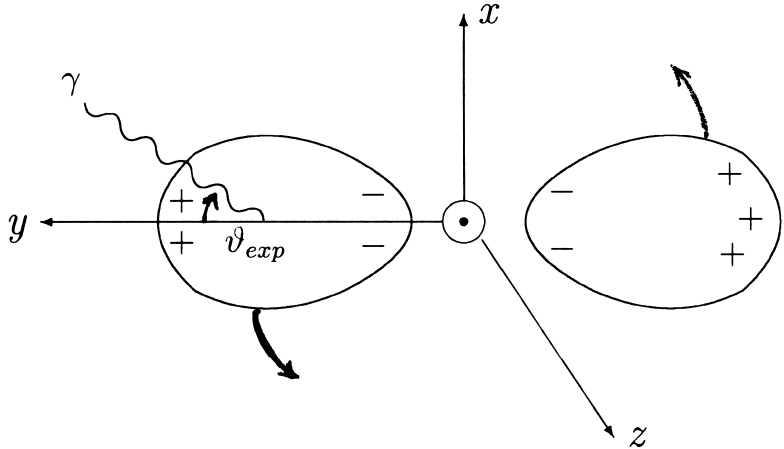}}
\caption{\footnotesize Scheme of study of the right-left asymmetry
in radiation from fission of $^{235}$U nucleus. Fission is induced
by a capture of a thermal neutron (designated by an open circle at
the origin) polarised along the $z$ axis, as shown by the point in
the middle. Pear-like fragments are polarised from the moment of
birth. Signs `$+$' and `$-$' show redistribution of charge inside
the fragments, which induces appearance of the electric dipole
moments. Circular arrows show the direction of rotation of the
fissile nucleus and nascent fragments caused by the polarisation
of the fissile nucleus. The fragments separate in the direction of
the $y$ axis; $\gamma$ quanta are registered in the ($x$, $y$)
plane at  the angle of $\vartheta_{exp}$ with respect to the $y$
axis. The probability of $\gamma$ emission at the given angle
changes when the direction of rotation is reversed together with
the neutron polarisation.  }
\end{figure}
Furthermore, the asymmetry parameter, $R(\vartheta_{exp})$ has been obtained.
That reads
\be
R(\vartheta_{exp}) = \frac {N_\gamma^\uparrow(\vartheta_{exp}) -
N_\gamma^\downarrow(\vartheta_{exp})}
{N_\gamma^\uparrow(\vartheta_{exp}) +N_\gamma^\downarrow(\vartheta_{exp})}\;,
\label{aseq}
\ee
where $N_\gamma^\uparrow$, $(N_\gamma^\downarrow)$ is the number of gammas
emitted at the given angle $\vartheta_{exp}$ with respect to the fission axis
when the neutron polarisation is along (right)
or counter (left)  the quantization axis $z$, respectively.
The experimental values reported in~\cite{dan} with 90\% longitudinally
polarised thermal neutrons are  \\
\hspace*{5em}$R^{exp}(35^\circ$) = (1.5 $\pm$ 0.4)$\times10^{-4}$, \\
\hspace*{5em}$R^{exp}(57^\circ$) = (2.3 $\pm$ 0.4)$\times10^{-4}$, and  \\
\hspace*{5em}$R^{exp}(90^\circ$) = ($-$0.2 $\pm$ 0.6)$\times10^{-4}$. \\
Earlier, similar effect has been found in $\alpha$ or another light charged
particle emission in ternary fission
\cite{ROT}. In  that case, the effect was attributed to the rotation of
the fission axis after the light charged particles are ejected.
Explanation of the effect needs that the alpha is emitted for the times of
the order of 10$^{-21}$ s, while the fragments did not separate
on a large distance yet.
Therefore, this effect is of great interest because it sheds light
on the dynamics of fission at a very early stage, also clarifying the
fundamentals of quantum mechanics.
As distinct from $\alpha$'s and other strongly interacting particles,
$\gamma$ emission mainly occurs from fully accelerated excited
fragments, past the neutron evaporation (e.g. \cite{D5}),
for characteristic times of 10$^{-14}$ -- 10$^{-12}$~s,
and therefore, cannot be explained as due to the same origin.

Our present purpose is to show that, on one side, the
Strutinsky---Denisov induced polarisation mechanism~\cite{D6,D7}
not only explains the non-statistical part of the spectrum, observed
in ref. \cite{ploeg}, but it is also strong enough, to explain
the results \cite{dan}.
In this case, the radiation is emitted from fragments
before the neutron emission due to snapping back the nuclear surface
within a time interval $\tau_{dis}$ which is determined by dissipation
of the collective energy. According to \cite{D5}, $\tau_{dis}\la 10^{-19}$~s.
Therefore, study of this effect gives invaluable information on
the dynamics of this process, providing a direct confirmation
of this phenomenon, which is of great interest, but very hardly
observed. Previous evidence of this effect was obtained in the shaking
muons emitted from the prompt fission fragments~\cite{book,D1,D2}.
On the other side, at the present stage of investigation, our calculation
suggests that the usual $\gamma$ radiation from fragments can be
expected to possess the same effect. However, even in this case, the
value of the asymmetry parameter (\ref{aseq}) depends on such a seemingly
intrinsic and hardly observable property as the angular velocity
of the spin rotation of the fragments, as well as dissipation
of the nuclear matter, providing a unique information on these primordial
features.

    Generally, right-left asymmetry can also be a manifestation
of violation of the space parity. Under space reflection $\m r \to -\m r$,
spin of the neutron, as a pseudovector, remains positively directed
along the new $z$ axis, but the right coordinate system goes over
the left one. Therefore, the asymmetry parameter (\ref{aseq})
changes the sign.    Whether parity violation was observed,
is a question of magnitude of the effect.
We return to this matter in section \ref{cncl}. We note that the
right-left asymmetry arises because the detecting system in \cite{dan}
does not distinguish between light and heavy fragments.
Otherwise, an effect of a triple correlation like
$(\m \sigma\cdot [\m p_f\times \m k_\gamma])$ could be observed,
which is $P$ even.
Therefore, one can use this circumstance in order to clarify the mechanism
of the observed phenomenon. Any effect of $P$ violation is ruled out by
mere fixing the direction of the heavy fragment and counting
the angle $\vartheta_{exp}$ off this direction.

   \section{Physical premises}

    First premise is alignment of the fission fragments.
Experiment shows that the fragments are formed partly aligned after scission
in the plane perpendicular to the fission axis, with the average spin of
$I$ = 7 -- 8. The effect of alignment was observed in the angular
distribution of the emitted radiation \cite{skars} and conversion muons
from the prompt fission fragments~\cite{np97}.
The spins of the fragments can be parallel or anti-parallel to each other.
In the last case, their total spin is compensated by the large angular
momentum of the relative motion of the fragments.
At the moment of scission, the nascent fragments have a pear-like form
with the noses directed towards the point of the rupture.
Wavefunction of a fragment can be presented as follows~\cite{BM}:
\be
\psi(\m r) = \left(\frac{2I+1}{16\pi^2}\right)^{1/2}
\begin{cases}
D^I_{MK}\phi_K(\m r') + (-1)^{I+K}D^I_{M\ -K}\phi_{-K}(\m r') &
\text{for $\pi$=1,}\\
i\left[D^I_{MK}\phi_K(\m r') -(-1)^{I+K}D^I_{M\ -K}\phi_{-K}(\m r')\right] &
\text{for $\pi$=-1.}
\end{cases}
\label{grg1}
\ee
$K$ is thus a good quantum number. Therefore,
evolution of the nuclear surface occurs under holding axial  symmetry
of the fragment. The symmetry axis rotates in the plane perpendicular
to the spin of the fissile nucleus~\cite{ka84}. We choose
the quantization axis $z$ along the neutron polarisation vector,
and the $y$ axis --- along the fragment direction.

    The second premise comes from the shake effects brought about
by the neck rupture. From mathematical view point,
the rupture means break-down of the analyticity of the Hamiltonian
with respect to time.  The rupture starts the snapping-back of the nuclear
surface \cite{halp}, which goes over the oscillations of the surface
smearing out in time. The lifetime of the oscillations is determined
by dissipation. It can be evaluated as
$\tau_{dis}\approx 10^{-19}$~s~\cite{D5}. The oscillations generate
electromagnetic field in space, changing with time. That causes
electromagnetic processes of internal conversion and $\gamma$ radiation.
Thus, muon shake in muon-induced
prompt fission manifests itself in muonic conversion.
The calculated probability agrees
with the experiment \cite{D1,D2}. In ref. \cite{dub93}, there was also
calculated the probability of emission of $\gamma$ quanta,
                   the results are also of interest in connection with experiments
\cite{ploeg,D4}, where the non-statistical $\gamma$ rays from $^{252}$Cf
spontaneous fission
are under investigation.  In the next section we revisit the results
\cite{dub93} in view of the present interest, make more detailed
calculations and correct misprints.
In section \ref{labsys}, we consider transformation of the angular
distribution from the intrinsic to the laboratory frame.
Numerical estimates are performed in
section \ref{num}. We summarise the results obtained in the conclusion
section, outline prospect for future research.

\section{Strutinsky---Denisov mechanism of pre-neutron emission of prompt
$\gamma$ rays}
\label{iedmnuc}

    The nuclear vibrations can be considered like the motion of a classical droplet.
Write down the conventional expansion of the nuclear form in spherical
harmonics
\be
R(\theta,\phi)=R_0\bigl(1+\beta_0+\sum_{\lambda,\mu}\beta_{\lambda,\mu}
Y_{\lambda,\mu}(\theta,\phi)\bigr)\;.
\label{Deq1}
\ee
Main properties of the fragments can be described with the allowance for the
quadrupole and octupole terms. This superposition of the even and odd
harmonics leads to a pear-like form of the nucleus. It is
essential that the electric dipole term in this case must be included
into expansion (\ref{Deq1})
to keep the centre of mass fixed \cite{D6}, the relation
$\beta_1= -0.743 \beta_2\beta_3$ following from the latter
condition~\cite{D1,D7}.
The other consequence is the appearance of the polarisation electric
dipole moment in the nucleus~\cite{D6}:
\be
d\equiv D/e = -\kappa \beta_2\beta_3\;. \label{Deq2}
\ee
The polarizability $\kappa$ in eq. (\ref{Deq2}) can be evaluated
e.g. from formulae~\cite{D6,D7}, which agree with experiment  (see also
other refs. in \cite{D7}).

Considering the oscillations quasiclassically and taking into account
the relaxation, put down
\be
\beta_i(t) = \beta_i^{(0)} \sin \omega_i t \, \exp (-\gamma_it/2)\;,
\quad i=2, \,3\;.       \label{Deq3}
\ee
Then the spectral density of the radiated energy is given in the classical
limit~\cite{lanIV}
by the following expression:
\be
d{\cal E}_\omega = \frac43 \bigl |\ddot D_\omega \bigr|^2\
\frac{d\omega}{2\pi}\;,    \label{Deq4}
\ee
where $D_\omega$ is the Fourier transform of the second derivative
of $D(t)$ with respect to time.
Using (\ref{Deq2}) and (\ref{Deq3}) in eq. (\ref{Deq4}), we find
\bea
\ddot D_\omega = -e\kappa \beta_2^{(0)} \beta_3^{(0)}
\int_0^\infty \exp(i\omega t)\frac{d^2}{dt^2}
\sin\omega_2t \sin\omega_3t \exp(-\gamma t/2)\ dt= \nonumber \\
= \frac i4 D_0 \frac{(\omega_2+\omega_3)^2}
{\omega-\omega_2-\omega_3+i\frac\gamma 2}\;,         \label{Deq5}
\eea
where $\gamma=\gamma_2+\gamma_3$ is the total quenching, and
$D_0\equiv ed_0=e\kappa \beta_2^{(0)} \beta_3^{(0)}$.

    Supposing  $\beta_2^{(0)}\approx\beta_3^{(0)}\approx 0.7$~\cite{D5},
we calculate by means of
formulae~\cite{D7} \mbox{$d_0\approx 5$~Fm}. Using then the LDM values
for a representative heavy fragment $^{140}$Xe,
which are \mbox{$\hbar \omega_2$ = 2.2 MeV},
\mbox{$\hbar \omega_3$ = 2.8 MeV}, and evaluating $\gamma$ from
the lifetime $\tau_{dis}=\gamma^{-1}=10^{-19}$~s,
as it is stated previously, and finally multiplying the result by a
factor of two in view of the presence of two fragments available,
one immediately finds by means of eq. (\ref{Deq4}) the number
of the electric dipole quanta
\be
N_d = \int_0^\infty \frac{d{\cal E}_\omega}{\hbar \omega}
\approx  \frac{e^2 d_0^2 \omega_0^3}{6\gamma} =
0.014\;\mbox{fission$^{-1}$}\;,     \label{Ng}
\ee
with $\omega_0 = \omega_2 + \omega_3 \approx$ 5~MeV being the resonance
frequency. We conclude that this value is in qualitative agreement with
experiment~\cite{ploeg,D4}, taking into account the uncertainties
connected with the value of $\tau_{dis}$.
For the value supposed, $\tau_{dis}\approx 10^{-19}$~s,
the contribution of the proposed mechanism is enough to explain the
experimental value.

    On the other hand, we see that this contribution is proportional to
$\tau_{dis}$.
Therefore, study of non-statistical $\gamma$ rays from fission gives
direct information about dissipation in large-amplitude collective
motion represented by postrupture oscillations in the fragments.

\section{Transformation to the laboratory system}
\label{labsys}

Let us start from consideration of classical radiation from a polarised
dipole system in the intrinsic coordinate system.
Then the angular distribution will be \cite{lanIV}
\be
\chi(\theta')= |Y_{11}(\theta', \phi')|^2 \sim \sin^2 (\theta')\;\;.
\label{grg2}
\ee
Distribution (\ref{grg2}) is normalised to the emission of one dipole
photon from the fragment. After the separation the axis of each fragment
rotates in the $(x, y)$ plane
with the  angular velocity $\omega$ (Fig. 1). The angle of the symmetry axis
at the moment of emission against the flight direction $y$ will be
\be
\beta(t) = \omega t\;.
\ee
Transformation from the intrinsic to the laboratory frame can be done
by means of two rotations at the Euler's angles $\beta$ against the axis $z$,
and then at $\pi/2$ against the new axis $y'$. The spherical functions
in (\ref{grg2}) in the laboratory system can be expressed in terms
of the generalised spherical Wigner functions  as follows:
\be
Y_{11} (\theta', \phi') =
\sum_m {\cal D}^1_{m1}\bigl(\beta(t), \frac\pi 2, 0\bigr)
Y_{1m}(\theta, \phi)\;.     \label{lab}
\ee
Inserting (\ref{lab}) into (\ref{grg2}) and using formulae \cite{edmon}
for the ${\cal D}$ functions, one arrives at the angular distribution
in the laboratory frame:
\be
\Chi(\theta,\phi; \beta(t)) =  \frac 3{8\pi}
\left[\sin^2\theta \cos^2(\phi-\beta) + \cos^2\theta\right]\;.
\ee
Integrating the last expression over time and taking into account
the time of relaxation due to dissipation of the collective energy
 $\tau_{dis}\equiv 1/\gamma$, we arrive at the angular distribution
in the laboratory system
\be
\Chi(\theta,\phi) =
\int_0^\infty \gamma e^{-\gamma t}\Chi(\theta,\phi; \beta(t)) \ dt =
\frac3{16\pi}\left[1 + \cos^2 \theta +
\sin^2\theta \cos 2\delta \cos 2(\phi-\delta)\right]  \;,      \label{grg6}
\ee
where
\be
\cos 2\delta = \frac{\gamma}{\sqrt{\gamma^2+4\omega^2}},\ \qquad
\sin 2\delta = \frac{2\omega}{\sqrt{\gamma^2+4\omega^2}} \;.
\ee
Introducing the anisotropy parameter $a$, $|a| \ll 1$,
we can put down instead of (\ref{grg6})
\be
\Chi(\theta,\phi) = 1 + a\sin^2\theta \cos 2\delta \cos 2(\phi-\delta) \;.
\label{grg6a}
\ee
Reversal of the polarisation of the incoming neutrons is
equivalent to the reversal of the direction of the rotation of the
fissile nucleus and the fragments. This leads to a replacement $\omega \ \to \ -
\omega$, that is $\delta \to -\delta$. Therefore, we arrive
by means of eq. (\ref{grg6}) at the following expression for
the asymmetry parameter:
\bea
R(\theta, \phi)= a\sin^2\theta\cos 2\delta\sin 2\delta\sin 2\phi =
  \frac12 a\sin^2\theta\sin 4\delta\sin 2\phi \;.
\label{grg7}
\eea

    Assuming the background radiation $\Chi_b(\theta, \phi)$ to be
isotropic, we can normalise it to the total number of the emitted photons
per fission $N_\gamma$ and arrive at the following expression:
\be
\Chi_b(\theta, \phi) = N_{\gamma} / 4\pi\;.
\label{param}
\ee
Given the number of the dipole quanta per fission $N_d$,  and
allowing for (\ref{grg6}),
one can put down an expression for the total angular distribution as follows:
\be
\Chi_b(\theta, \phi) + \Chi(\theta, \phi) = \frac1{4\pi}N_{\gamma}
+ \frac3{16\pi}N_d \sin^2\theta \ \cos 2\delta \ \cos 2(\phi-\delta)  \;.
\label{anis}
\ee
Comparing (\ref{anis}) with (\ref{grg6a}), we arrive at the following
expression for the anisotropy parameter $a$:
\be
a = \frac34 \frac{N_d}{N_\gamma}\;.   \label{par}
\ee

    The above expression (\ref{grg7}) for the asymmetry parameter $R$ is remarkable
in many respects.
First, it is anti-symmetric with respect to the value of $\phi = \pi / 2$.
The asymmetry parameter increases from zero to maximal value for
$0\le\phi < \pi/4$.
Then it diminishes again to  zero, changing the sign at $\phi = \pi/2$,
and becomes negative for  $\pi / 2<\phi < \pi$.
Then it varies in the same manner in the open angle of
$\pi \le \phi < 2\pi$.  This behaviour is similar to
the previously reported angular distribution which is characteristic
to the ROT effect in ternary fission \cite{ROT}.

    Second, that vanishes in the two opposite limiting cases:
of very slow and very fast rotation as compared to the time of dissipation.
Note that  for the purpose of comparison with experiment~\cite{dan},
making use of (\ref{theta}), the angular distribution in the laboratory
system (\ref{grg6}) for $\theta = \pi/2$ can be rewritten in the form
\be
\Chi(\frac\pi2,\phi) = \frac3{16\pi} \left[
1-\cos2\delta\cos2(\vartheta_{exp}+\delta)\right] =
 \frac3{16\pi} \left[
1-\cos^2\delta + 2\cos^2\delta\sin^2 (\vartheta_{exp} + \delta) \right] \;,
\label{grg6b}
\ee
which is similar to (\ref{grg2}),
but with shifted by angle $\delta$ argument, plus an isotropic term.
Correspondingly, the asymmetry parameter (\ref{grg7}) becomes
\be
R = \frac12 a \sin 4\delta \sin 2\vartheta_{exp}  \;.
\label{grg7a}
\ee
In the first case $\omega \ll \gamma$, the shifting angle remains small.
It reads:
\be
\delta \approx \frac{\omega}  \gamma\;.     \label{slow}
\ee
(\ref{slow}) has a transparent physical sense of the mean fraction
of a full revolution the fragment performs
before radiating. The optimal value is achieved with $\sin 4\delta$ = 1,
or $\delta = \pi/8 = 22.5^\circ$. In this case,
\be
R_{max} = \frac12 a \sin 2\vartheta_{exp}\;.
\label{grg7b}
\ee

    At first sight, similar mechanism
was considered  in \cite{ROT} for explanation of the observed
shift of the peak of the angular distribution of the ternary light particle,
depending on the direction of polarisation of the incoming neutron,
with rotation of the
fission axis instead of rotation of the symmetry axis of a fragment.
Eq. (\ref{slow}) gives a scale of the shift.
However, the nature of the shift in our case is different.
In ref. \cite{ROT}, the fission axis rotated at a small angle of about
one degree after the emission of the ternary particle in the direction which
was determined by the direction of polarisation of the fissile nucleus.
Probably, that was the reason why the observed effect was called ROT effect.
This effect should be considerably damped by the Coulomb interaction
of all the three particles in the final state.
The interaction is absent in the case of $\gamma$ emission, therefore,
the emitted $\gamma$'s bare information on the prompt orientation
of the fragments at  the moment of emission. As distinct to the emission
of the light charged particles, we neglect rotation of the inter-fragment
axis, which is small in comparison with rotation of the fragments
(see section \ref{num}).

    In the second case $\omega \gg \gamma$, that is the fragments
revolve many times before emission. As a result,
the effect of asymmetry averages and smoothes out. This is manifested
in the fact that the value of
$\delta \to \frac\pi 4$\ , and the asymmetry parameter (\ref{grg7}),
(\ref{grg7a}) vanishes together with $\sin 4\delta$.
This case can be applied to the usual radiation from fragments which
occurs after neutron evaporation within characteristic times of
$\sim 10^{-14} - 10^{-12}$~s.
For these $\gamma$ quanta, the right-left effect \emph{vanishes},
in spite of that the angular distribution of the radiation
qualitatively remains similar to (\ref{grg2}) \cite{stru}.

\section{The results and discussion}
\label{num}

    First, we note that the experimental data \cite{dan} exhibit
the angular dependence of the asymmetry parameter similar to
(\ref{grg7a}): the experimental value changes the sign when $\vartheta_{exp}$
crosses $\pi / 2$. Regarding the numerical values, simple
solid-body estimation shows that with spin $I$ = 1, the fragments
revolve with $\omega \approx 2.1\times 10^{19}$~s$^{-1}$.
Therefore, with $\gamma \approx$ 10$^{19}$~s$^{-1}$ one obtains
$\cos 2\delta \approx$ 0.24.

    Next task is evaluation of the anisotropy parameter $a$ in
(\ref{par}). The $N_d$ value was calculated in section \ref{iedmnuc}.
Let us consider the radiation background $N_b$. As a result of fission,
deformed primary fragments are obtained with the excitation energy of
around 12 MeV. The excitation energy is partly fallen down
with the evaporated neutrons. Already the neutrons are emitted by the
fully accelerated fragments, being kinematically shifted by the
translation velocity of the fragments \cite{D5}. Only recently it was
established that there is a certain fraction of prompt neutrons
at the level of  10\% which are emitted from the fission area before
the fragments are accelerated.
    There is much less known about prompt gamma rays.
The nonstatistical part of the spectrum \cite{ploeg} can be of this kind,
as shown in section \ref{iedmnuc}. These gamma quanta can contribute
to the effect of the right-left asymmentry, as we saw in the above section.
Brehmstrahlung gamma rays also can be emitted during acceleration
of the fragments.
Recently, brehmstrahlung was discovered in alpha decay (e.g. \cite{brink}).
Angular distribution of the  brehmstrahlung quanta would be the same
as that in (\ref{grg2}), and therefore, that  would be accompanied
by a related right-left effect due to rotation of the inter-fragment axis.
However, fission fragments are much heavier than alphas, and moreover,
the dipole moment of a pair of fragments is close to zero because of
near constancy of the fragment $Z/A$ ratio.
For these reasons, one should not expect any essential manifestation of
the brehmstrahlung. Estimates can be done similar to those for the induced
electric dipole radiation which were made in section \ref{labsys}.
Let $Z_1$, $A_1$, $Z_2$ and $A_2$ be the atomic and mass numbers
of the heavy and light fragments, respectively, and $Z = Z_1+Z_2$,
$A=A_1+A_2$ are the atomic and mass numbers of the fissile nucleus,
respectively.
The energy radiated during time $dt$ reads as follows \cite{lanIV}:
\be
dI_\gamma / dt = \frac23\text{\it\"d}\ ^2\;.            \label{bs1}
\ee
In the center of mass system of the fragments the electric dipole moment is
\be
d/e = (Z_1A_2-Z_2A_1) R / A =
\left[ (Z_1^0+q)A_2 - (Z_2^0-q) A_1 \right] R/A
= Rq\;,               \label{bs2}
\ee
where $Z_i^0 = A_i (Z/A)$, $i = 1, 2$, and $R$ is the interfragment distance.
Therefore, effective charge of the fragments for the brehmstrahlung is
\be
q = Z_i-Z_i^0\;, \qquad |q|  \ll Z_i\;.
\ee
On the other hand, from the Newton's law
we get for the acceleration of the fragments
\be
\text{\it\"R} \equiv \text{\it\"d}/qe = \nu/MR^2\;, \label{newtn}
\ee
with $\nu = Z_1Z_2e^2$, and $M$ being the reduced mass of the fragments.
Substituting (\ref{newtn}) into (\ref{bs1}) and integrating over
all the fission trajectory from the initial point $R_0$, which we will
fix by assuming zero total kinetic energy of the fragments at this point,
to infinity, one gets         the expression for the radiated energy:
\be
I_\gamma = (qe)^2\frac{2\sqrt{2}\sqrt{E-\nu x}}{45\nu M^{3/2}}
 \left[ 3(E-\nu x)^2 - 10E (E-\nu x) +15E^2\right]
\Bigr|_{1/R_0}^0 =
\frac{16\sqrt 2}{45}\frac{q^2 E^{5/2}}{Z_1Z_2  M^{3/2}}\;,
\label{bst}
\ee
with  $E$ being the total kinetic energy of the separated fragments.

    For a given mass number, fragment distribution over $Z$ is close to
the Gauss distribution with the dispersion $\sigma \approx$ 2. With this
characteristic value of $q$ and mean total kinetic energy $E = 160$ MeV,
by means of eq. (\ref{bst}) one finds
\be
I_\gamma = 2.0\cdot10^{-5} \text{ MeV}\;.
\label{bsn}
\ee
An estimate for the number of brehmstrahlung quanta with the energy
within a domain of 100~keV to 5~MeV
\be
10^{-5}\text{ fission}^{-1}\leq N_\gamma \leq
2\cdot10^{-4}\text{  fission}^{-1}
\label{bsnf}
\ee
follows (\ref{bsn}). As expected, these values are too low to produce
a noticeable effect of right-left asymmetry.

    According to the experimental spectrum \cite{peelle},
average number of quanta emitted per fission is
$N_\gamma\approx 8$~fission$^{-1}$. Assuming the isotropic
angular distribution of these quanta,
and given the total probability of the electric dipole quanta of (\ref{Ng})
with the angular distribution (\ref{grg2}), we can estimate the  parameter
$a$ to be $a \approx \frac34 \cdot 0.014 / \ 8 \approx 0.0013$.
By means of (\ref{grg7a}) we then  get an estimation of the
asymmetry parameter
$R(\vartheta_{exp})= 2.8\times 10^{-4}$ for the angles of
$\vartheta_{exp}$ = 35$^\circ$ and 57$^\circ$, and $R(90^\circ)$ = 0.
These values are close to the experimental values~\cite{dan} for these
angles cited in Introduction. Note that for a slower rotation
of the fragments, the value of $\cos 2\delta$ could be higher. The maximal
value of the asymmetry parameter (\ref{grg7b}) could be by a factor
of 2 higher.

    Up to now, however, we did not discuss the energy dependence of the effect.
In the above example, we retained only the resonant term in (\ref{Ng}),
corresponding to emission of the quanta
with the total energy of approximately 5 MeV.  From the experimental spectrum
from fission \cite{peelle} one can conclude that effect-to-background ratio
would be much better in this domain. Thus, there is only $\sim 0.3 $
quanta per fission with the energy $E_\gamma >$ 2.5~MeV.  With this background
value, one arrives at the asymmetry parameter
$a = \frac34\cdot 0.014/0.3$ = 0.035, and  the corresponding
right-left effect $R$ at the level of 10$^{-2}$.

\section{Conclusion}
\label{cncl}

1. It is shown that the observed effect of right-left asymmetry
in gamma radiation can be
explained as due to emission from rotating fragments. The rotation
is due to the
primary polarisation of the fissile nucleus whose rotational moment
partly transfers to the fragments after the scission.

2. According to (\ref{Ng}), this effect is predicted for the $\gamma$ quanta
of sufficiently high energy, approaching the giant dipole resonance.
This is in accordance with the time scale, which is of the order of
10$^{-19}$~s. For smaller energies, the emission probability is lower
by orders of magnitude. The radiative lifetime is less than one period of rotation of a fragment. Therefore, study of the effect for such hard gammas
presents an information about the process of fission at this early stage.

3. For further understanding, the energy dependence of the effect is
highly needed to be measured in experiment.    In the above example,
we retained only the resonant term corresponding to emission of the quanta
with the total energy of approximately 5~MeV. We neglected the other resonant
term with the difference energy in (\ref{Ng})
$\omega_{dif}=\omega_3-\omega_2$ = 0.6 MeV,
because in this case the expected emission probability
is by three orders of magnitude
less, being proportional to the cube of the energy.
An important conclusion for experiment follows this result.
In order to study early stages of fission, one should detect
hard $\gamma$ quanta with the energy in the region of giant dipole
resonance (though reduced considerably by elongation of the fissile
nucleus), as in refs. \cite{ploeg,peelle}.

4. The mechanism discussed above has nothing in common with
parity nonconservation. Manifestation of the latter in fission generally
starts at the level of  10$^{-4}$.
Violation of conservation of parity remains the next candidate
to be at least partly responsible for the observed result. Therefore,
more detailed experiments should shed light on the origin of the effect.
However, equally in this case, the $\gamma$'s must be emitted
within the time interval of about $\sim 2\pi/\omega\sim10^{-19}$~s, as
shown in section \ref{labsys}. Moreover, abstracting from the background,
one can conclude from (\ref{grg6}) that the right-left effect by itself
in prompt $\gamma$ rays would be of the order of 10$^{-1}$, which magnitude
is much higher than that which could be expected
from the parity violation effects.

5. Estimates are made of the brehmstrahlung energy from fragments.
The estimates show that the brehmstrahlung in fission is very weak,
in contrast with that in $\alpha$ decay.

6. Study of the above effect gives information on the dynamics of fission
at the stage of snapping back the nuclear surface of the born fragments.
Hence that is  a direct confirmation
of this phenomenon~\cite{halp}, which is of great interest, but very hardly
observed. Previous evidence of this effect was obtained in the shaking
muons emitted from the prompt fission fragments~\cite{book,D1,D2}.

\bigskip
The author is grateful to G.V.Danilyan for inducing discussions.
He would also like to thank F.G\"onnenwein for reading the manuscript
and many helpful remarks.


\newpage


\begin{thebibliography}{99}


\bibitem{dan} G.V.Danilyan. Invited talk at the XLI PNPI Winter School
    on the Nuclear and Particle Physics, St. Petersburg, Repino,
    19 -- 24 February 2007; ISINN---2008, June 11-14, 2008, Dubna, Russia..
\bibitem{ROT} F.Goennenwein, M.Mutterer, A.Gagarski, I.Guseva,
    G. Petrov, et al., Phys. Lett. {\bf B652} (2007) 13.
\bibitem{D6} V.M.Strutinsky. At. Energ., 1956, No. 4, p. 150. ({\it In Russian.})
\bibitem{D7} V.Yu.Denisov. Yad. Fiz.  {\bf 55} (1992) 2647;
     {\bf 49} (1989) 644.
\bibitem{ploeg} H. van der Ploeg et al. Phys. Rev. Lett. {\bf 68} (1992) 3145;
    KVI annual report, 1991, p. 17.
\bibitem{D5} Yu. P. Gangrsky, B. N. Markov, V. P. Perelygin. Registratsia
    i sipektrometria oskolkov delenia. Moscow: Energoizdat, 1992
    (in Russian) (Registration and Spectrometry of the fission fragments)
\bibitem{book} F.F.Karpeshin. Fission in Muonic Atoms and Resonance
    Conversion. St.-Petersburg: ``Nauka'', 2006. ({\it In Russian.})
\bibitem{D1} F.F.Karpeshin. Z. Phys. A{\bf 344} (1992) 55; Yad. Fiz.
    {\bf 55} (1992) 2893.
\bibitem{D2} G. Ye.Belovitsky et al. In: ``Fiftieth Anniversary or Nuclear
    Fission'', Proc. Intern. Conf., St. Petersburg, 1989. V. 1, P. 313.
\bibitem{skars} Skarsvag K.  Phys. Rev. {\bf C22} (1980)  638.
\bibitem{np97}  F.F.Karpeshin.  Nucl. Phys. {\bf A617} (1997) 211.
\bibitem{BM} A.Bohr and B.Mottelson. Nuclear Structure. Vol. II.
    W.A.Benjamin, Inc. New York, Amsterdam, 1974.
\bibitem{ka84} F.F.Karpeshin. Yad. Fiz. {\bf40} (1984) 643.
    ({\it Engl. transl.} Sov. J. Nucl. Phys. (USA), {\it 40} (1984) 412.)
\bibitem {halp} J. Halpern.  Ann. Rev. Nucl. Sci. {\bf 21} (1971)  245.
\bibitem{dub93} F.F.Karpeshin.  In: International School-Seminar
    on Heavy Ion Physics.  Ed.: Yu.Ts.Oganessian, Yu.E.Penionzhkevich
    and R.Kalpakchieva. Dubna: 1993. Vol. 1, P. 294; arXiv:0710.1743v1.
\bibitem{D4} A.Wiswesser e.a. GSI annual report, 1991, P. 79.
\bibitem{lanIV} L.D.Landau, E.M. Lifshitz. Teoria Polya, Moscow: Nauka, 1973.
    ({\it In Russian}) (Theory of Field)
\bibitem{edmon}  A.R.Edmonds. Angular moments in Quantum Mechanics.
    CERN 55-26, 1955.
\bibitem{brink} D.M.Brink. In: Fission Dynamics of Atomic Clusters and
    Nuclei. Ed. by J. da Providencia, D.M.Brink, F.Karpechine
    and F.B.Malik. World Scientific, New Jersey--- London---
    Singapore---Hong Kong, 2001, P. 248.
\bibitem {peelle} R.W. Peelle, F.S. Meienshein. Phys. Rev.
    {\bf C3} (1971) 373.
\bibitem{stru} V.M.Strutinsky. ZhETP,  {\bf 38} (1959) 861.

\end{thebibliography}
\end{document}